\documentclass{PoS}
\usepackage{graphicx}

\makeatletter
\newsavebox{\@parc@ption}
\def\parcaption#1{%
\sbox{\@parc@ption}{\shortstack[l]{#1}}%
\setbox\@tempboxa\hbox{\csname fnum@\@captype\endcsname}%
\@tempdima\columnwidth \advance\@tempdima-\wd\@tempboxa
\@tempdimb.8\@tempdima 
\ifdim\wd\@parc@ption>\@tempdimb \@tempdima\@tempdimb
\else\@tempdima\wd\@parc@ption\fi
\sbox{\@tempboxa}{\parbox[t]{\@tempdima}{#1}}%
\caption{\usebox{\@tempboxa}}}
\makeatother

\title{Leibniz rule, locality and supersymmetry on lattice}

\ShortTitle{Leibniz rule, locality and supersymmetry on lattice}

\author{Mitsuhiro Kato\\
 Institute of Physics, University of Tokyo, Komaba, Meguro-ku, Tokyo 153-8902, Japan \\
        E-mail: \email{kato@hep1.c.u-tokyo.ac.jp}}

\author{Makoto Sakamoto\\
        Department of Physics, Kobe University, Nada-ku, Hyogo 657-8501, Japan\\
        E-mail: \email{dragon@kobe-u.ac.jp}}

\author{\speaker{Hiroto So}%
         \thanks{
                  This work was supported in part by
          Grants-in-Aid for Scientific Research ( No.20340048, No.20540274 and No.22540281) 
 by the Japanese Ministry of Education, Science, Sports and Culture.}\\
           Department of Physics,  Ehime University, Bunkyou-chou 2-5, 
 Matsuyama 790-8577, Japan\\
        E-mail: \email{so@phys.sci.ehime-u.ac.jp}}

\abstract{
In a finite volume system, we prove a no-go theorem on a Leibniz rule  
with a care of   locality argument on latttice.  
The new possibility  on the Leibniz rule solutions on lattice is discussed. 
 Although the new solution  admits  a local difference operator, 
a non-local product rule is needed. In the case,  a supersymmetric interacting theory  is simply realized.   
The difference between finite flavor systems and matrix representations of  infinite flavor systems 
is explained based on a finite volume system analysis including the no-go theorem.

}
\FullConference{The XXX International Symposium on Lattice Filed Theory\\
		 June 24-29,2012\\
		 Cairns,  Australia}

\begin{document}

\section{Motivations}
When we construct  interacting supersymmetric theories  on lattice, we must take care of 
a Leibniz rule on lattice. A  no-go theorem on the rule for an infinite volume  system  on lattice
  has been proved by us \cite{no-go}. 
In its  proof, there are two important clues, i.e.  
  translational  invariance  and   locality for an infinite  system. 
 The holomorphic function property associated with a lattice operator
   can decsribe those both clues. 
 To manipulate the locality in a finite volume system, 
 we must seek  another discrete version of holomorphism which expresses  translational invariance and  locality. 
 Corresponding to a certain translational invariant operator,  a discrete function  instead of a  complex function  can be defined  
  and we can describe it as  a local operator in the finite volume system.

  There is a puzzling situation between a multi-flavor system and  the matrix representation of an infinite flavor system which  
  matrix product and a commutator difference operator  is satisfied with the Leibniz rule \cite{dublin}.  
  To the contrary,
  there is a Leibniz rule no-go theorem  on lattice for the finite flavor case. Cannot we recognize a matrix representaion 
  as inifinite flavor number limit of a multi-flavor system? 
  In order to solve the problem, we must analyze  
  the finite number case of the flavor $N_f$ and the lattice size  $N$ which implies the spatial volume.

\section{Definition of a local lattice theory}
In this article, we treat a one-dimensional system. The extention to higher dimensions 
can be realized by the direct product of higher  dimensional coordinates. 
We shall start  with a setup for a local lattice theory in a finite system. 
The system size and  lattice constant  are denoted  as $N$ and $a_L=1$ , respectively.  
 We  impose a   the periodic boundary condition,  $\phi_n=\phi_{n+N}$ on any lattice field.

A general difference operator on lattice is defined  as

\begin{equation}
(D\phi )_n  \equiv \sum\limits_m^N  {D_{nm} } \phi _m ,
\end{equation}
\noindent
where $\sum_m^N D_{nm}=0$ due to a vanishing  constant mode.
For a product rule, we generally define as 
\begin{equation}
(\phi  \times \eta )_n  \equiv \sum\limits_{m\ell }^N  {C_{nm\ell }^{} } \phi _m \eta _\ell .
\end{equation}
\noindent
In the next step, we concentrate on translational invariant  theories from general lattice ones
\footnote {The translational invariance strongly connects with the momentum conservation law.  
The strange momentum conservation on lattice can be realized as \cite{Dondi-Nicolai,Kawamoto}. }. 
 As the result,  the difference operator and the product rule have the following property, 
\begin{equation}
D_{n\,\,m}^{}  = D_{n + k\,\,\,m + k}^{}  = D(n - m)=D(n-m+N)
\end{equation}
\noindent
and 

\begin{equation}
C_{nm\ell }^{}  = C_{n + k\,\,\,m + k\,\,\,\ell  + k}^{}  = C(n - \ell ,m - \ell ) 
=C(n - \ell+N ,m - \ell )=C(n - \ell ,m - \ell +N)  ,
\end{equation}
\noindent
where the periodic boundary condition is used.  
By $N$-root of unity,

\begin{equation}
w_a  \equiv e^{\frac{{2\pi ia}}{N}}  = \omega _N^a ,\,\,\,z_b  \equiv e^{\frac{{2\pi ib}}{N}}  
= \omega _N^b ,\,\,\,\omega _N  
\equiv e^{\frac{{2\pi i}}{N}} ,\,\,\,\omega _N^N  = 1,
\end{equation}
\noindent
we define  "momentum"  representations and their complex extentions,

\begin{eqnarray}
\hat D_a  &\equiv& \sum\limits_m^N {w_a^m } D(m) = \sum\limits_m^N {\omega _N^{am} } D(m) , \nonumber \\
  \hat D_{a + i\varepsilon N}  &\equiv& \sum\limits_m^N {\omega _N^{(a + i\varepsilon N)m} } D(m) , \nonumber\\
 \hat C_{ab} & \equiv &\sum\limits_{m,n}^N {w_a^m z_b^n C(n,m)}  = \sum\limits_{m,n}^N {\omega _N^{am + bn} C(n,m)} ,  \nonumber\\
\hat C_{a + i\varepsilon N,b + i\eta N} & \equiv & \sum\limits_{m,n}^N {\omega _N^{(a + i\varepsilon N)m + (b + i\eta N)n} C(n,m)} ,
\end{eqnarray}
\noindent
of the difference operator and the product rule 
where we must note that indices $a,b$ are discrete momentum labels. 
The locality in a finite volume system is defined as 
\begin{equation}
\left| {\,D\left( { - n} \right)\,} \right| \le K\,e_{}^{ - \kappa \left| n \right|} \,\,\,,\,\,\,\kappa 
 > {\rm{0}}\,,\,\,\,{\rm{for}}\,\,\,\,1 <  < |n|\, <  < \,N  ,
\end{equation}
\noindent
for large $N$. 

Using the orthogonality and the completeness property,

\begin{equation}
\sum\limits_a^N {\omega _N^{(n + m)a} }  = N\delta _{n + m,0} ,\,\,\,\,\,\sum\limits_n^N {\omega _N^{(a + b)n} }  = N\delta _{a + b,0} ,
\end{equation}
\noindent
the sufficient condition for  locality in the finite system is that there exists nonzero finite $\varepsilon$ with 

\begin{equation}
\label{locality-condition}
\mathop {{\rm{max}}}\limits_a \left\{ {\,\,\left| {\hat D_{a + i\varepsilon N} } \right|\,} \right\} = O(N^0 ),\,\,\,\,\,\,\mathop {{\rm{max}}}\limits_{a,b} \left\{ {\,\,\left| {\hat C_{a + i\varepsilon N,b + i\eta N} } \right|\,} \right\} = O(N^0 ) ,
\end{equation}
\noindent
because 
\begin{eqnarray}
 \frac{1}{N}\sum\limits_a^N {\omega _N^{n(a + i\varepsilon N)} \hat D_{a + i\varepsilon N}  = } 
 \frac{1}{N}\sum\limits_a^N {\omega _N^{an} \hat D_a  = } D( - n)_{} , \nonumber \\
\left| {D\left( { - n} \right)\,} \right| \le \frac{1}{N}\sum\limits_a^N {\left| {\hat D_{a + i\varepsilon N} } \right|}
 \,\,e_{}^{ - 2\pi \varepsilon |n|}  \le \mathop {{\rm{max}}}\limits_a \left\{ {\,\,\left| {\hat D_{a + i\varepsilon N} }
  \right|\,} \right\}\,e_{}^{ - 2\pi \,\varepsilon \,|n|}  , 
\end{eqnarray}
\noindent
where the first equality is  similar to the Cauchy's integral theorem about a complex function owing to 
independence on $\varepsilon$.  
For the rule $C$, there are similar inequalities.  
On the other hand, 
in the case of 
\begin{equation}
\left| {\,D\left( { - n} \right)\,} \right| \le K\,e_{}^{ - \kappa \left| n \right|} \,\,\,,\,\,\,\kappa  > {\rm{0}}\,,\,\,\,{\rm{for}}\,\,\,\,1 <  < |n|\, <  < \,N  ,
\end{equation}
\noindent
  with
\begin{equation}
\kappa  \pm 2\pi \varepsilon  > 0 ,
\end{equation}
\noindent
its necessary condition can be verified as 
\begin{equation}
\left| {\,\hat D_{a + i\varepsilon N} \,} \right| = \left| {\sum\limits_m^N {\omega _N^{m(a + i\varepsilon N)} D\left( m \right)\,} } \right| \le \sum\limits_m^N {e^{ - 2\pi \varepsilon m} } \left| {D\left( m \right)\,} \right| \le K\sum\limits_m^N {e^{ - \left( {\kappa  \pm 2\pi \varepsilon } \right)\left| m \right|} }  = O\left( {N^0 } \right) .
\end{equation}
\noindent
For product rule, similarily, in the case of 

\begin{equation}
\left| {\,C\left( { - m, - n} \right)\,} \right| \le K\,e_{}^{ - \kappa \left| m \right| 
- \lambda \left| n \right|} \,\,\,,\,\,\,\,\,\kappa  > 0,\,\,\,\lambda  > 0, 
\end{equation}
\noindent
with
\begin{equation}
\kappa  \pm 2\pi \varepsilon  > 0,\lambda  \pm 2\pi \eta  > 0,
\end{equation}
\noindent
the necessary condition can be verified as 
\begin{eqnarray}
\left| {\hat C_{a + i\varepsilon N,b + i\eta N} } \right| &=& 
\left| {\sum\limits_{m,n}^N {\omega _N^{m\left( {a + i\varepsilon N} \right) 
+ n\left( {b + i\eta N} \right)} C\left( {m,n} \right)} } \right| \nonumber \\
& \le& \sum\limits_{m,n}^N {e^{ - 2\pi \left( {\varepsilon m + \eta n} \right)} } 
\left| {C\left( {m,n} \right)} \right| \nonumber \\
 &\le& K\sum\limits_{m,n}^N {e^{ - \left( {\kappa  \pm 2\pi \varepsilon } \right)\left| m \right| 
 - \left( {\lambda  \pm 2\pi \eta } \right)\left| n \right|} }  = O\left( {N^0 } \right) .
\end{eqnarray}

In the summary of this section, we have proposed that (\ref{locality-condition}) is 
the necessary and sufficient conditions for the locality in a finite volume system.

\section{Finite size no-go theorem for Leibniz rule on lattice}

A no-go theorem states that
 translation invariance,  locality, a Leibniz rule and nontrivial product  
      cannot  be simultaneously   satisfied on a finite volume lattice. 
The  Leibniz rule by using  only  translation invariance condition can be rewritten as 
\begin{equation}
\label{Leibniz-1}
\hat C_{a,b} (\hat D_{a + b}  - \hat D_a  - \hat D_b ) = 0 .
\end{equation}
\noindent
For any $a$ and $b$, if  $\hat C_{a,b}  \ne 0$,  then (\ref{Leibniz-1}) says 
\begin{equation}
\label{Leibniz-2}
\hat D_{a + b}  - \hat D_a  - \hat D_b  = 0 .
\end{equation}
\noindent
The general solution of (\ref{Leibniz-2})is given by
\begin{equation}
\label{SLAC-solution}
\hat D_a  = \frac{a}{b}\hat D_b  \propto a  .
\end{equation}
\noindent
The solution (\ref{SLAC-solution}) is SLAC-type \cite{SLAC} owing to  
\begin{equation}
\omega _N^a  = e^{\frac{{2\pi ia}}{N}}  = e^{ip} \,\, \Rightarrow  a \propto p ,
\end{equation}
\noindent
and is non-local because of 
\begin{equation}
\mathop {\max }\limits_a \left\{ {\left| {\hat D_{a + i\varepsilon \,N} } \right|} \right\} 
\ge \frac{1}{N}\sum\limits_a^N {\left| {\,\hat D_{a + i\varepsilon \,N} \,} \right|} 
 = \frac{1}{N}\sum\limits_a^N {\left| {\,a + i\varepsilon N\,} \right|} 
 \left| {\,\frac{{\hat D_b }}{b}\,} \right| = O\left( {N\,^{\rm{1}} } \right) \ne O(N^0 )  .
\end{equation}
\noindent
If  $\hat C_{a,b} $
       is  nontrivial, 
        then    $D(n)$  is  SLAC-type and it is nonlocal.  q.e.d

One of   possible solutions may be the case that $D$ is local but $C$ is trivial or nonlocal. 
We can illustrate one example: ${\hat C_{a,b}} = K{\delta _{a + b,1}}$  .
If $K=O(N^0)$, then the product rule is trivial which leads us to  a trivial continuum limit. 
 On the other hand, if $K=O(N^1)$, then it is just nonlocal.   
The real space expression is 

\begin{equation}
{C_{lmn}} = C(l - m,l - n) = \frac{K}{N}{\delta _{m,n}}\omega _N^{ - (l - n)}  .
\end{equation}
\noindent
Its  Leibniz rue realization is the followings,

\begin{equation}
{\hat D_1} = {\hat D_a} + {\hat D_{1 - a}}  .
\end{equation}
\noindent
This relation  is not difficult to construct its local solutions. 
Using one of these solutions, we can write an explicit supersymmetric interacting action,  

\begin{eqnarray}
S = \frac{1}{2}D\phi  \cdot D\phi  + i\bar \psi  \cdot D\psi  + F \cdot F + \frac{{ig}}{2}F \cdot \left( {\phi  \times \phi } \right) + ig\phi  \cdot \left( {\bar \psi  \times \psi } \right)\\
\phi  \cdot \chi  \equiv \sum\limits_n^N {{\phi _n}{\chi _n},\,\,\,\,\,} {\left( {\phi  \times \chi } \right)_n} \equiv \sum\limits_{m,l}^N {{C_{nml}}{\phi _m}{\chi _l}\,\,\,\,\,} 
\end{eqnarray}
\noindent
where the supersymmetry can be defined as 
\begin{eqnarray}
\delta \phi & =& \varepsilon \bar \psi  + \psi \bar \varepsilon , \nonumber\\
\delta \psi & =& \varepsilon \left( {iD\phi  + F} \right),\,\,\delta \bar \psi  = \bar \varepsilon \left( { - iD\phi  + F} \right), \nonumber\\
\delta F &=&  - i\varepsilon D\bar \psi  - i\bar \varepsilon D\psi  .
\end{eqnarray}

\section{Multi-flavor system  and matrix representation}

In a finite multi-flavor system on lattice, no-go theorem on a Leibniz rule   
    can be proved   \cite{no-go}    but 
 there exists the Lebniz rule through a matrix product rule in the matrix representation of an  infinite flavor system \cite{dublin}. 
This apparent inconsistency or the curious flavor  limit can be solved by classifying 
two kinds of flavors after  appropriate flavor diagonalization.

For a finite flavor  $N_f$ and a finite volume $N$ system,  
product rule and a difference operator are  defined  as  

\begin{equation}
\left( {\varphi  \times \chi } \right)_l^p  \equiv \sum\limits_{m,n,q,r}^{N,N_f } {C_{lmn}^{pqr} } 
\varphi _m^q \chi _n^r ,\,\,\,\,\left( {D\varphi } \right)_l^p  \equiv \sum\limits_{m,q}^{N,N_f } {D_{lm}^{pq} }
\end{equation}
\noindent
where indices $p,q,r$ are flavor ones.
From these translation invariance,  we introduce the following notation,  
\begin{equation}
C_{lmn}^{pqr}  = C^{pqr} (m - l,n - l),\,\,\,D_{lm}^{pq}  = D^{pq} (m - l). 
\end{equation}
\noindent
The $w$-representations($N$-root expressions) of the product rule and the difference operator   are expressed as 
\begin{equation}
\hat C_{LM}^{pqr}  \equiv \sum\limits_{m,n}^N {\omega _N^{Lm + Mn} C^{pqr} (m,n)} 
,\,\,\,\hat D_L^{pq}  \equiv \sum\limits_m^N {\omega _N^{Lm} } D^{pq} (m) .
\end{equation}
\noindent
Their flavor matrix forms are 
\begin{equation}
\left( {\hat C_{LM}^q } \right)_{pr}  \equiv \hat C_{LM}^{pqr} ,\,\,\,\left( {\hat D_L^{} } \right)_{pq}  \equiv \hat D_L^{pq} .
\end{equation}
\noindent
The 
Jordan's standard  form  of $\hat{D}$
leads us to the following parametrization in the flavor matrix, 
\begin{equation}
\hat{D} \to U\hat{D}U^{ - 1}  = \hat{D}_{diag}  + \hat{E}_+  ,
\end{equation}
\noindent
where 
\begin{equation}
\hat{D}_{diag}^{pq}  = \hat \Delta ^p _L \delta _{pq} ,\,\,\,\,\hat E_ + ^{pq}  
= \varepsilon _L^p \delta _{p,q + 1} ,\,\,\,\,\,\,\hat \Delta ^p _0  = 0,\,\,\,\,\varepsilon _L^p  = 0~{\rm or~}1  .
\end{equation}
\noindent
Then, we can write the Leibniz rule by a flavor matrix form

\begin{equation}
\Big( \hat{\Delta}_{L+M}^p -   \hat{\Delta}_{L}^q-\hat{\Delta}_{M}^r  \Big)\hat C_{L,M}^{pqr} 
=\hat R_{L,M}^{pqr} (\epsilon) ,
\end{equation}
\noindent
where 
\begin{equation}
\hat R_{L,M}^{pqr} (\varepsilon ) \equiv  - \varepsilon _{L + M}^p \hat C_{L,M}^{p - 1~qr} 
 + \varepsilon _L^{q + 1} \hat C_{L,M}^{p~q + 1~r}  + \varepsilon _M^{r + 1} \hat C_{L,M}^{pqr + 1}  .
\end{equation}
\noindent
For a finite flavor system, since 

\begin{equation}
1 \le p,q,r \le N_f, ~~\hat C_{L,M}^{0qr}=\hat C_{L,M}^{p~N_f+1~r}=\hat C_{L,M}^{pq~N_f+1}=0,    
\end{equation}
\noindent
it follows as 
\begin{equation}
\label{R-equation}
R_{L,M}^{pqr} (\varepsilon )=0 .
\end{equation}
\noindent
Therefore, the Leibniz rule of a finite flavor by a flavor matrix form is rewritten as 
\begin{equation}
\left( {\hat \Delta _{L + M}^p  - \hat \Delta _L^q  - \hat \Delta _M^r } \right)\hat C_{L,M}^{pqr}  
\equiv \hat D_{L,M}^{pqr} \hat C_{L,M}^{pqr}  = 0. 
\end{equation}
\noindent
The solution is easily found as 
\begin{equation}
\hat D_{L,M}^{pqr}  = 0\,\,\,\,\,{\rm{or}}\,\,\,\,\hat C_{L,M}^{pqr}  = 0 .
\end{equation}
\noindent
$\hat D_{L,M}^{pqr}  = 0$ case leads us to $\hat \Delta _{}^p (w_a) = 0$ 
in the local lattice theory framework, 
using  a finite  system no-go theorem which is proved in the previous section. 
Consequently, for a finite-flavored infinite volume system, 
we have the following two kinds of flavors;  flavor type-A    means  
that it has a trivial   difference operator, $\hat \Delta^p (w) = 0$ and flavor type-B 
does  a  trivial  field product    $\hat C_{L,M}^{pqr}  = 0 $ between  its flavors.

The next stage is the analysis for  the matrix representation  of an infinite flavor and infinite volume system. 
In the representation, we treat field variables, 
$\Phi_{ij}$ 
where both $i$ and $j$ run from $1$ to $N_{\rm matrix}$.
We must consider an  $N=N_f=2N_{\rm matrix}$  case   because 
the following identification is realized,
\begin{equation}
\Phi_{ij}= \phi_{n=(i+j)}^{p=(i-j)} .
\end{equation}
\noindent
The product rule between   matrices  leads us to  

\begin{equation}
C_{lmn}^{pqr}  \propto \delta _{p,q + r} \delta _{n - l,q} \delta _{l - m,r}  \,  ,
\end{equation}
\noindent
and the commutator difference operator corresponds to 

\begin{equation}
\,[d,~\Phi]_{ij} = (D\phi)_{n=(i+j)}^{p=(i-j)}        ,
\end{equation}
\noindent
where $d$ implies some anti-hermitian $N_{\rm matrix} \times N_{\rm matrix}$ matrix. 
For this  matrix representation, since we impose the usual periodic boundary condition for 
both a lattice space and a flavor space,  the following relation  
    
\begin{equation}
 R_{L,M}^{pqr} (\varepsilon )\ne 0 ,
\end{equation}
\noindent
is generated  inevitably.  
This relation  is the essential  difference for usual finite flavor systems. 
Furthermore, owing to $R\ne 0$, there is always mixing  between coupling-free flavor-B and 
motion-free flavor-A.

\section{Summaries}

We have proved a Leibniz rule no-go theorem in 
a finite volume system.  
Instead of holomorphic functions for infinite volume systems, we used 
the discrete bounded functions. Then, we can classify cases keeping the rule on lattice 
into  the following three ones: 

1~     If we take a local nontrivial product $C_{lmn}$, then  
      $D_{mn}$ is always SLAC-type~(nonlocal).

2~     If $D_{mn}$ is local, then $C_{lmn}$ is nonlocal or trivial. 
New possibility  supersymmetry application with the strange momentum conservation law~\cite{Dondi-Nicolai,Kawamoto}.
 
3~       
      If $D_{mn}$  is SLAC-type~(nonlocal), then $ C_{lmn} $ is arbitrary.

\noindent
In the case of the second possibility, we can construct an explicit supersymmetric action with interactions.

For a finite flavor system versus the matrix representation in the infinite flavor system ,  we make a table: 

\begin{center}
\begin{tabular}{|c|c|c|c|c|} \hline
 & the number of & Leibniz rule & locality& A-B separation \\ 
  &  components &  &  &   \\ \hline 
 multi flavor & $N_f \times N$& no-go & local & yes\\ \hline
 multi flavor & $N_f \times N$& no-go & nonlocal &yes  \\ \hline
 matrix representation&  $N \times N=N_{\rm matrix} \times N_{\rm matrix}$& escape &nonlocal by $N$ infinity & no\\ \hline
\end{tabular}
\end{center}


\begin{thebibliography}{99}
\bibitem{no-go} M.~Kato, M.~Sakamoto, and H.~So, 
\emph{Taming the Leibniz Rule on the Lattice},
{\bf ~JHEP} 0805(2008) 057.
\bibitem{dublin} M. Kato, M. Sakamoto and H. So, 
\emph{Leibniz rule and exact supersymmetry on lattice: 
A Case of supersymmetrical quantum mechanics}, 
{\bf ~PoS LAT2005} (2006) 274.
\bibitem{Dondi-Nicolai}  P.H. Dondi and H. Nicolai, 
\emph{Lattice Supersymmetry},
{\bf ~Nuovo Cim.~A41~}(1977)~1. 
\bibitem{Kawamoto} A.~D'Adda, I.~Kanamori, N.~Kawamoto and  J.~ Saito
\emph{Species Doublers as Super Multiplets in Lattice Supersymmetry: 
Chiral Conditions of Wess-Zumino Model for D=N=2},
{\bf ~JHEP} 0805(2008) 057.
\bibitem{SLAC} S.D. Drell, M. Weinstein and S. Yankielowicz, 
\emph{Strong-coupling field theories. II. 
Fermions and gauge fields on a lattice}, 
{\bf ~Phys. Rev. D14}{~(1976)~}{1627}.

\end{thebibliography}
\end{document}